\documentclass[twocolumn,english,prl,superscriptaddress,showpacs]{revtex4-1}
\usepackage[T1]{fontenc}
\usepackage[latin9]{inputenc}
\usepackage{amstext}
\usepackage{graphicx}

\makeatletter
\usepackage[colorlinks,citecolor=darkblue,linkcolor=darkred,urlcolor=darkblue] {hyperref}
\usepackage{xcolor}
\definecolor{darkblue}{rgb}{0.1,0.2,0.6} \definecolor{darkred}{rgb}{0.8,0.1,0.2}

\makeatother

\usepackage{babel}
\begin{document}


\title{Many-Body Localization in System with a Completely Delocalized Single-Particle
Spectrum}

\author{Yevgeny Bar Lev}
\email{yb2296@columbia.edu}

\author{David R. Reichman}

\affiliation{Department of Chemistry, Columbia University, 3000 Broadway, New
York, New York 10027, USA}

\author{Yoav Sagi}

\affiliation{Department of Physics, Technion \textendash{} Israel Institute of
Technology, Haifa 32000, Israel}
\begin{abstract}
Many-body localization (MBL) in a one-dimensional Fermi Hubbard model
with random on-site interactions is studied. While for this model
all single-particle states are trivially delocalized, it is shown
that for sufficiently strong disordered interactions the model is
many-body localized. It is therefore argued that MBL does not necessary
rely on localization of the single-particle spectrum. This model provides
a convenient platform to study pure MBL phenomenology, since Anderson
localization in this model does not exist. By examining various forms
of the interaction term a dramatic effect of symmetries on charge
transport is demonstrated. A possible realization in a cold atom experiments
is proposed.
\end{abstract}
\maketitle
\emph{Introduction}.\textendash It has been known for almost 60 years
that \emph{non-interacting} particles in one-dimensional disordered
systems exhibit Anderson localization \cite{Anderson1958b}. Transport
in these systems is exponentially suppressed with the system size,
and without coupling to the environment, these systems are non-ergodic
at any temperature. Anderson localized systems are, however, non-generic,
since they do not include interactions which allow for the exchange
of energy. For many years it was assumed that interactions generally
restore ergodicity and destroy localization \cite{Fleishman1978}.
A decade ago, using non-equilibrium diagrammatic techniques, it was
argued that Anderson localization is stable under the addition of
a small short-ranged interactions \cite{Basko2006a}, a phenomenon
currently known as many-body localization (MBL). Many-body localized
systems are the only known \emph{generic} non-ergodic systems which
do not follow the assumptions of statistical mechanics \cite{Altman2014,Nandkishore2014,Vasseur2016}.
While the realization of MBL systems presents challenges in condensed
matter systems due to inevitable presence of phonons \cite{Basko2007a,Ovadia2014},
recent experiments in cold atoms have provided evidence of the existence
of MBL in both one-dimensional \cite{Schreiber2015a,Bordia2015,Smith2015}
and two-dimensional systems \cite{Choi2016}.

To establish the existence of MBL, the seminal work of Basko, Aleiner
and Altshuler assumes the presence of quenched disorder and localization
of \emph{all} single-particle states \cite{Basko2006a}. It is currently
under debate whether quenched disorder is necessary for the existence
of MBL. A number of numerical studies of \emph{translationally invariant}
systems with no quenched disorder have been carried out. However,
due to large finite size effects these studies are inconclusive with
respect to localization \cite{Schiulaz2013,Grover2013,Schiulaz2014,Yao2014,Hickey2014,Papic2015,Horssen2015,Pino2015}.
A related question, whether MBL can exist in a system where only \emph{some}
of the single-particle states are delocalized, namely in systems with
a mobility edge in the single-particle spectrum, has been affirmatively
answered \cite{Li2015,Li2016,Modak,Modak2015}. In our work, we go
one step beyond, and completely abolish the assumption of localization
of single-particle states. We show that many-body localization is
possible when the non-interacting limit is trivial, namely when \emph{all}
single-particle states are completely delocalized. A related result
has been discussed from a different perspective in a recent study
of the XXZ model (see Appendix of Ref.~\cite{Vasseur2015b}). While
previously studied MBL systems could be viewed as continuous deformations
of the Anderson insulator \cite{Bauer2013,Imbrie2014}, much in the
same vein as a Fermi liquid is a continuous deformation of a Fermi
gas, our work suggests that there are distinct classes of systems
exhibiting MBL that differ in their global symmetries.
\begin{figure}
\begin{centering}
\includegraphics[width=86mm]{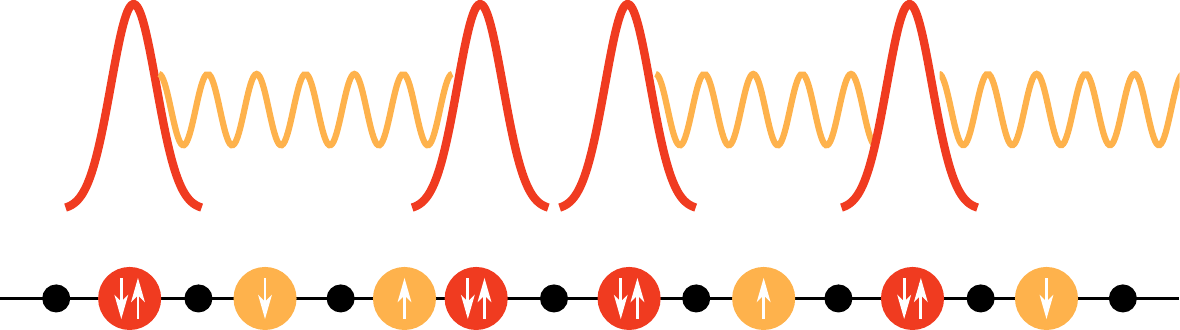}
\par\end{centering}
\caption{\label{fig:model-sketch}A schematic of the model considered in this
work. The lower portion of the figure shows a random configuration
of particles. The red (dark) particles are doublons, and the orange
(light) particles are singlons. The upper portion of the figure is
a cartoon of the charge distribution, with localized doublons and
delocalized singlons.}
\end{figure}

\emph{Model}.\textendash We study the dynamical properties of one-dimensional
Fermi Hubbard model with a random interaction term, 
\begin{equation}
\hat{H}_{ns}=-t_{h}\sum_{\sigma,i=1}^{L-1}\left(\hat{c}_{i\sigma}^{\dagger}\hat{c}_{i+1,\sigma}+\hat{c}_{i+1,\sigma}^{\dagger}\hat{c}_{i,\sigma}\right)+\sum_{i=1}^{L}U_{i}\hat{n}_{\uparrow i}\hat{n}_{\downarrow i},\label{eq:ham_non_sym}
\end{equation}
where $L$ is the length of the lattice, $\hat{c}{}_{i\sigma}^{\dagger}$
$\left(\hat{c}_{i\sigma}\right)$ is the creation (annihilation) operator
of site $i$ and spin $\sigma$ obeying the usual anti-commutation
relations, $\hat{n}_{i\sigma}=\hat{c}_{i\sigma}^{\dagger}\hat{c}_{i\sigma}$
is the number operator, $t_{h}$ is the hopping strength, which we
will set to one, the interaction terms $U_{i}$ are random and uniformly
distributed on the interval $-\Delta U\leq U_{i}\leq0$. We use a
definite (attractive) sign of the interaction, but for the infinite
temperature limit considered here, we have verified that the sign
of the interaction does not change the conclusions of our work. The
single particle states of this model are simple plane waves and therefore
without interactions this model is trivially delocalized and thus
cannot be studied by the perturbation theory developed in Ref.~\cite{Basko2006a}.
It also cannot be studied by the local unitary diagonalization technique
of Ref.~\cite{Imbrie2014}, since the starting diagonal Hamiltonian
is highly degenerate. Nevertheless, some intuition can be acquired
by considering the hopping term as a perturbation. At the lowest non-trivial
order in the hopping, the model effectively contains two species (see
Fig.~\ref{fig:model-sketch}): doubly charge excitations - doublons,
and singly charge excitations - singlons. The singlons are light and
hop at a rate, $t_{h}$, while the doublons are heavy and hop at the
average rate $\sim4t_{h}^{2}/\Delta U$. Since the interaction is
random, the doublons are strongly localized, while, the singlons do
not see an effective disordered potential, and thus are delocalized.
Nevertheless, for any initial state with a finite doublon density,
the doublons serve as ``random barriers'' to the singlons, which
leads to their localization. 

\emph{Results}.\textendash To verify that over time the singlons do
not delocalize the doublons we use numerically exact methods: exact
diagonalization (ED) and time-dependent density matrix renormalization
group (tDMRG) \cite{Vidal2003}. We calculate the spread of a charge
excitation at infinite temperature, by evaluating the correlation
function, 
\begin{equation}
C_{i}\left(t\right)=\frac{1}{\text{Tr }\hat{P}}\text{Tr }\hat{P}\left(\hat{n}_{i}\left(t\right)-1\right)\left(\hat{n}_{0}-1\right),
\end{equation}
where $\hat{P}$ is a projector, which we define as $\hat{P}_{s}$
(no doublons), $\hat{P}_{d}$ (no singlons) and $\hat{I}$ (infinite
temperature), and $\hat{n}_{i}\equiv\hat{n}_{i\uparrow}+\hat{n}_{i\downarrow},$
measures the total charge (or total number of atoms, in the case of
neutral ultracold atoms) at site $i$. Since we aim to demonstrate
localization, we fix the disordered interaction to be large enough
(see Ref.~\cite{SuppMat2016}), $\Delta U=30$, and leave the exploration
of transport \emph{across} the MBL transition for a subsequent work.
To characterize the spread of the charge excitation we calculate its
width as a function of time,
\begin{equation}
\sigma^{2}\left(t\right)=\sum_{i}i^{2}C_{i}\left(t\right),
\end{equation}
and average it over random initial configurations of the particles
as well as the disordered interaction. To eliminate finite size effects,
we make sure that the excitation has not reached the boundaries of
the system for the simulation times, which is achieved by correspondingly
increasing the size of the system. Thus the dynamics we calculate
correspond to the bulk limit up to all times observed. 
\begin{figure}
\begin{centering}
\includegraphics[width=86mm]{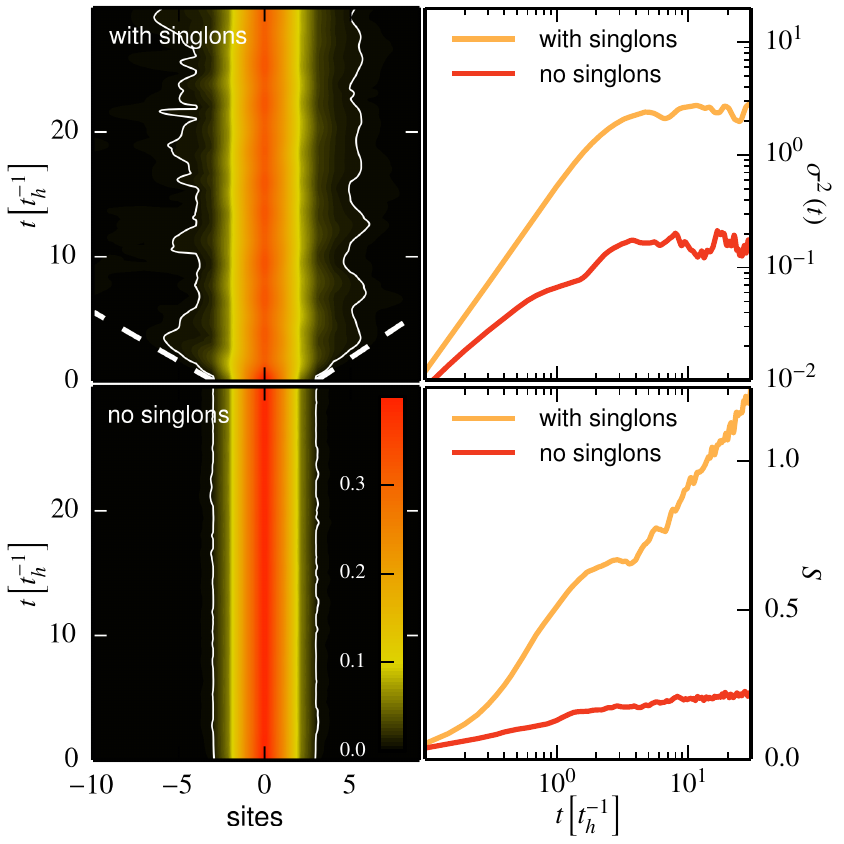}
\par\end{centering}
\caption{\label{fig:ns-singlons-v-non-singlons}Charge excitation dynamics
of the random interaction Fermi-Hubbard model (\ref{eq:ham_non_sym})
with and without singlons in the initial state. The left panels display
contour plots of the correlation function, $C_{i}\left(t\right)$,
as a function of space and time. The solid white line corresponds
to the contour line, $C_{i}\left(t\right)=7.5\times10^{-3}$, and
the dashed white lines highlight ballistic jets of singlons. The top
right panel presents the width of the excitation as a function of
time on a log-log plot, and the bottom right panel the entanglement
entropy $S$ as a function of time on a semi-log plot. The data has
been obtained using tDMRG for $L=20$ and $\Delta U=30$, and averaged
over a minimum of 300 realizations.}
\end{figure}
In Fig.~\ref{fig:ns-singlons-v-non-singlons} we present results
of tDMRG simulations for a system of size $L=20$ at an average filling
of 0.5. For this simulation we have used a discarded weight of $\chi=10^{-9}$,
a second order Trotter decomposition, and a step size of $\delta t=0.05$.
We have ensured that within statistical error the result is converged
with respect to the time step and the discarded weight. On the left
side of Fig.~\ref{fig:ns-singlons-v-non-singlons} we compare the
spread of charge excitation starting from two different initial conditions.
On the bottom panel we exclude singlons $\left(\hat{P}=\hat{P}_{d}\right)$
from the initial random configurations of charges, while on the top
panel singlons are not excluded $\left(\hat{P}=\hat{I}\right)$. With
singlons the excitation has initial ballistic jets which disappear
on length scales longer than the mean free path of the singlons, which
for infinite temperatures considered here, corresponds to the average
distance between the blocking doublons ($l_{e}\approx4$). As can
be seen from the bottom left panel, the ballistic jets vanish after
singlons are removed from the initial configurations. For both initial
conditions, the width of the excitation of the charge initially grows
ballistically, but then saturates to a finite plateau value, indicating
localization (right top panel). The entanglement entropy growths logarithmically,
which is typical for many-body localized systems \cite{Znidaric2008,Bardarson2012}.
For random initial charge configurations \emph{without} doublons the
excitation appears delocalized for timescales on which bulk transport
is accessible (data not shown). While the putative delocalization
of this initial condition could be a result of a mobility edge, the
strength of the disordered interaction was chosen such that \emph{all}
many-body states are localized, namely there is \emph{no} mobility
edge (see Supp. Matt. \cite{SuppMat2016}). After a short time (see
next paragraph) a finite density of doublons of the order of $O\left(1/U\right)$
will be generated, which will result in eventual localization of the
singlons. The expected localization length should be at least of the
order of the distance between the doublons, $\lambda_{s}\approx U=30$,
and is beyond the system sizes and times available in our simulations.
The dramatic difference in short time dynamics, highlights the importance
of proper selection of initial conditions for cold atom experiments.
While the system is localized for a \emph{typical} initial condition
(as we see from Fig.~\ref{fig:ns-singlons-v-non-singlons}), the
initial configurations without doublons can show putative delocalization
for quite long times. Although these states are of measure zero in
the thermodynamic limit, they are still realizable in cold atom experiments,
where the density of doublons or singlons can be effectively controlled
\cite{Schreiber2015a}.

There are three different simple time-scales in model (\ref{eq:ham_non_sym}):
$t_{s}=t_{h}^{-1}$, which corresponds to hopping of the singlons,
$t_{d}\sim U/\left(4t_{h}^{2}\right)$, which corresponds to hopping
of the doublons and for temperatures, $T\ll U$, there is a time scale
which corresponds to the decay (generation) of the doublons. This
timescale can be formidably long, $t_{\text{decay}}\sim\exp\left(cU\right)$
(where $c$ is some constant) \cite{Strohmaier2010,Sensarma2010},
however, for infinite temperatures studied here, thermal fluctuations
provide the necessary energy to break the doublon apart, such that
doublons decay occurs at the timescale of $t_{s}$. Therefore the
longest time-scale in our problem is $t_{d}$. To verify that the
observed localization exists also for times much larger than this
timescale namely, $t\gg t_{d}\approx7$, we utilize exact diagonalization.
\begin{figure}
\begin{centering}
\includegraphics[width=86mm]{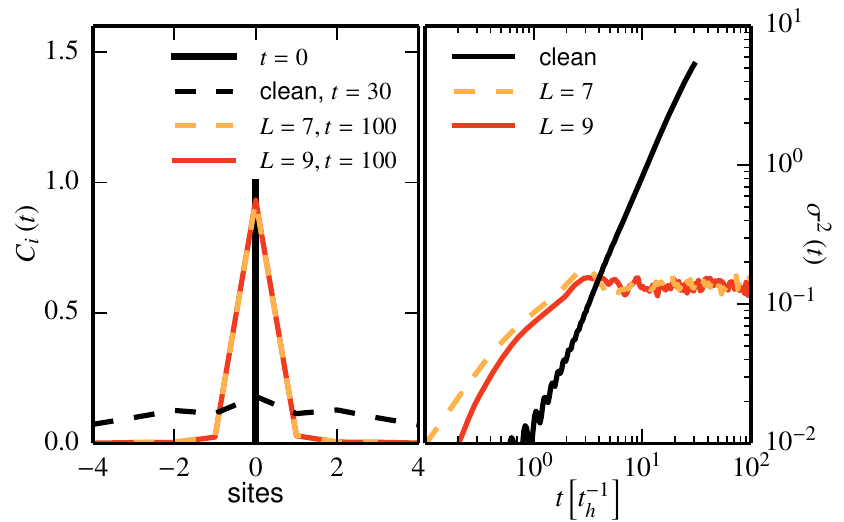}
\par\end{centering}
\caption{\label{fig:ed-clean-vs-disorder}Charge excitation dynamics for clean
and disordered Hubbard models starting from a random charge configuration
without singlons. \emph{Right panel}: charge excitation width as a
function of time for disordered colored (gray) lines and clean Hubbard
models (black dashed line) on a log-log plot. The parameters used
for the disordered case are $-30\leq U_{i}\leq0$, and for the clean
case $U_{i}=-30$. \emph{Left panel}: Excitation profile for the initial
(full black line) and final times (gray, and dashed black lines).}
\end{figure}
As is clear from Fig.~\ref{fig:ns-singlons-v-non-singlons}, without
the singlons the excitation is effectively contained in a region of
less than 10 sites. We therefore limit our exact diagonalization simulations
to system sizes $L=7$ and $L=9$. In Fig.~\ref{fig:ed-clean-vs-disorder}
we show the width of the excitation as a function of time up to time
$t=100$, starting from an initial state without singlons. Clearly,
localization persists up to this time, and finite size effects are
negligible. This can be also inferred from the profile of the excitation
at the final time of the simulation, which lies away from the boundaries
(see left panel). We can also compare to the dynamics in the \emph{clean}
case, with same interaction strength, $U_{i}=-30$. In this case,
the model can be effectively described by the Heisenberg model \cite{Chao1978}.
Over the same timescale for which localization persists in the \emph{disordered}
system, in the \emph{clean} system the excitation rapidly spreads
over the entire lattice (see left panel). The entanglement entropy
spreads ballistically (not shown), and bulk charge transport (before
the excitation has reached the boundaries) is super-diffusive, $\sigma^{2}\left(t\right)\propto t^{1.65}$,
consistent with previous studies \cite{Znidaric2011}\footnote{Due to the small size of our system, the obtained exponent is somehow
larger that the exponent obtained in Ref.~\cite{Znidaric2011}, which
is $4/3$. }. We therefore conclude that the observed localization is \emph{not}
related to the slow drift of the doublons over timescale, $t_{d}$,
but is true many-body localization. 

After establishing localization for the model (\ref{eq:ham_non_sym})
we consider the effect of symmetries on non-equilibrium dynamics.
We add an additional $SU(2)$ symmetry in the charge sector by changing
the interaction term (see Supp. Matt. \cite{SuppMat2016}),
\begin{equation}
\sum_{i}U_{i}\hat{n}_{\uparrow i}\hat{n}_{\downarrow i}\to\sum_{i}U_{i}\left(\hat{n}_{\uparrow i}-\frac{1}{2}\right)\left(\hat{n}_{\downarrow i}-\frac{1}{2}\right).\label{eq:interaction_variants}
\end{equation}
For a spatially independent interaction, $U_{i}=U$, this change corresponds
to a shift in the chemical potential, which leaves the non-equilibrium
dynamics unaffected. This is however \emph{not} the case for a spatially
dependent interaction, where the additional symmetry dramatically
affects the dynamics. Naively, by expanding the RHS of (\ref{eq:interaction_variants})
we obtain an effective disordered potential, $\sum_{i}U_{i}\hat{n}_{i}/2$,
which might lead one to suspect that the system is localized (note
that the single-particle spectrum is still trivial). This reasoning
is however misleading, since the potential and the disordered interaction
are perfectly correlated and therefore a more detailed analysis is
in order. If the hopping term is set to zero, the eigenstates of the
system including the ground state are highly degenerate, since moving
a doublon to an empty site does not cost energy. Therefore the doublons
do not ``feel'' the presence of an effective disordered potential.
In the limit of \emph{zero} hopping the system is trivially localized;
for small, but \emph{non-zero} hopping, $\Delta U\gg t_{h}$, and
for initial condition without singlons, the dynamics of the system
is effectively described by the random Heisenberg model,

\begin{equation}
\hat{H}_{\text{eff}}=\sum_{ij}\frac{2t_{h}^{2}}{\left|U_{ij}\right|}\left(\hat{\mathbf{S}}_{i}\cdot\hat{\mathbf{S}}_{j}-\frac{1}{4}\right),\label{eq:random_heisenberg}
\end{equation}
where the $\hat{\mathbf{S}}_{i}$ are spin-1/2 operators, $U_{ij}\equiv\left(U_{i}+U_{j}\right)/2$,
and the derivation was performed along the lines of the derivation
of the Heisenberg model from the Hubbard model \cite{Chao1978,SuppMat2016}.
The random Heisenberg model was previously studied using strong disorder
renormalization group. While the ground state is a localized random
singlet state \cite{Ma1979}, at finite temperatures the renormalization
group breaks down, which previously was interpreted as the onset of
delocalization \cite{Agarwal2015,Vasseur2015}. Our results are consistent
with this prediction. 
\begin{figure}
\begin{centering}
\includegraphics[width=86mm]{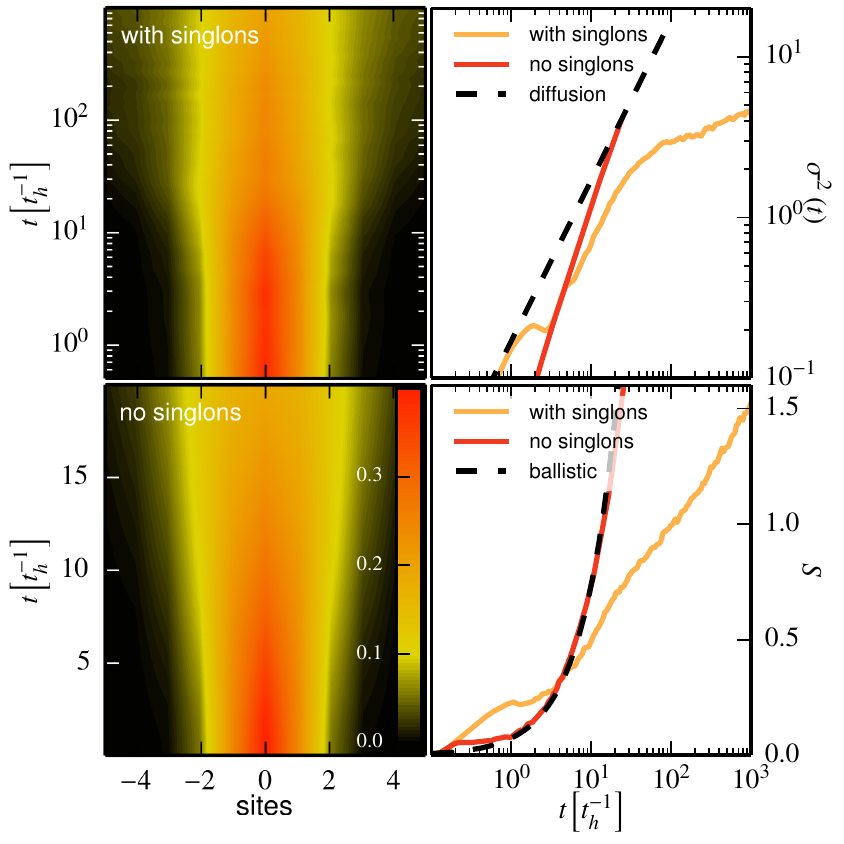}
\par\end{centering}
\caption{\label{fig:s-singlons-v-non-singlons} Same as Fig.~\ref{fig:ns-singlons-v-non-singlons}
but with interaction term of Eq.~(\ref{eq:interaction_variants}).
The data has been obtained using ED for $L=11$ and $\Delta U=30$,
and averaged over at least 300 realizations. Note that the simulation
time was $t=1000$ for initial conditions with singlons, and $t=100$
for the initial conditions without singlons.}
\end{figure}
By projecting away all the singlons from the random initial configurations,
we use ED to study transport \footnote{Interestingly, the times for which bulk transport can be obtained
in ED are very similar to the times accessible in tDMRG for same numerical
cost. We therefore use ED for this part. }. We find (see Fig.~\ref{fig:s-singlons-v-non-singlons}) that entanglement
entropy spreads ballistically, and charge excitations propagate super-diffusively.
However, the super-diffusive propagation of the charge is likely a
result of the relatively short times for which bulk transport is accessible
in our simulations $\left(t\approx25\right)$, and the asymptotic
charge transport is probably diffusive. Indeed for domain wall initial
conditions we were able to observe diffusion even on this short timescale
\cite{SuppMat2016}. By performing a canonical transformation, $\hat{c}_{i\uparrow}\to\hat{c}_{i\uparrow}^{\dagger},$
$\hat{c}_{i\downarrow}\to\left(-1\right)^{i}\hat{c}_{i\downarrow}^{\dagger}$,
the Hamiltonian maps to $-H,$ namely, the many-body spectrum of this
model is symmetric with respect to zero energy. Moreover, this transformation
maps doublons and holons into singlons, which suggests that if doublons
are delocalized (as we have shown), also singlons are delocalized
for the transformed problem. Since the dynamics under $-H$ is equivalent
to dynamics under $H$ with a reversed direction of time, and since
$H$ is time-reversal invariant, we conclude that singlons are delocalized
for $H$ itself. Interestingly, starting with an initial condition
which includes all possible charge configurations, namely a mix of
doublons, holons and singlons, renders the charge transport \emph{slower}.
After a a relatively short diffusive regime, transport becomes sub-diffusive,
or perhaps even logarithmic. Correspondingly, the entanglement entropy
crosses-over from ballistic growth to a growth which is slightly faster
than logarithmic (see Fig.~\ref{fig:s-singlons-v-non-singlons}).
The mechanism behind the observed slow charge transport is currently
not clear, and more detailed consideration of finite size effects
in this regime is needed. It is however clear that the seemingly minor
change in the form of the interaction (\ref{eq:interaction_variants})
dramatically changes the system dynamics and leads to delocalization.

\emph{Experimental implementation}.\textendash The Fermi-Hubbard model
has been extensively studied with ultracold atoms \cite{Esslinger2010}.
Tight-binding is achieved by loading the atoms to the lowest band
of an optical lattice and the strength of interaction is controlled
by tuning the s-wave scattering length using a magnetic Fano-Feshbach
resonance. We suggest here to implement spatially random interactions
between particles by means of optical Feshbach resonance with a random
optical control. With the recent advances in quantum gas microscopy,
this will allow the scattering strength to be controlled on a sub-micron
spatial resolution. Optical Feshbach resonances are known to incur
excess heating due to spontaneous emission from the excited state,
and that could be detrimental for realizing many-body localization.
To mitigate this effect, we suggest to use a scheme in which the light
couples the bound Feshbach molecular state to an excited molecular
state off-resonantly \cite{Bauer2009,Bauer2009a}. Using this scheme
the heating time can be as long as $10$ms \cite{Fu2013}, which is
about $50t_{h}^{-1}$ \cite{Schreiber2015a}. Since our numerical
results show that localization occurs in less than $10t_{h}^{-1}$
we conclude that losses incurred by the optical control technique
should not prevent one from observing and detecting the MBL phase.

\emph{Discussion}.\textendash{} We have established many-body localization
in one-dimensional random Hubbard model with a completely delocalized
single-particle spectrum which is \emph{not} amenable to the theoretical
analysis of Refs.~\cite{Basko2006a,Imbrie2014}. We proposed a realization
of this model in cold atom experiments using spatially resolved optical
Feshbach resonances \cite{Bauer2009,Bauer2009a}. In this model, many-body
localization follows from fragmentation of particles into slow and
fast species (doublons and singlons), which is a result of the strong
interactions. One species is Anderson localized by the quenched disordered
potential and localizes the other species by creating an effective
chain of randomly distributed barriers. The mechanism for localization
that we posit here is similar to recent proposals for MBL in clean
systems, where two different species are introduced from the outset
\cite{Grover2013,Schiulaz2013,Schiulaz2014,Yao2014,Papic2015,Pino2015,Antipov2016},
with the crucial difference that the effective disorder in our case
is quenched, while it is annealed in the clean case. We believe that
this is the main reason why, unlike here, the observation of MBL is
challenging in the clean case \cite{Papic2015}, unless one of the
species is completely immobile \cite{Antipov2016}. We demonstrated
the importance of the initial conditions for observation of MBL in
experimentally attainable timescales, and have shown that localization
is absent when the model has\emph{ an additional }$SU\left(2\right)$
symmetry in the charge sector.

\emph{Note} \emph{added}.\textendash{} During the completion of the
manuscript three studies have appeared with relevance to the work
presented here. Ref.~\cite{Sierant2016} presents a numerical study
of a distinct model with a completely delocalized single particle
spectrum. In Ref.~\cite{Garrison2016} a related \emph{translationally
invariant} model, which is also $SU\left(2\right)$ symmetric, was
numerically studied. For sufficiently high interaction strength and
generic initial conditions the authors show evidence of nonergodic
behavior, consistent with our observations here. The work of Ref.~\cite{Potter2016}
advocates for the impossibility of MBL in a system with non-Abelian
continuous symmetry, which is \emph{inconsistent} with our results.
For this reason we present an extensive analysis of possible finite
time and size effects in the Supplementary Materials \cite{SuppMat2016}.
It remains unclear if all arguments of Ref.~\cite{Potter2016} apply
to our model. In particular, we work with initial charge configurations
which \emph{individually} dynamically break the $SU\left(2\right)$
symmetry, such that the symmetry is only satisfied after an average
over \emph{all} charge configuration is performed.
\begin{acknowledgments}
\emph{Acknowledgemnts}.\textendash{} YBL would like to thank Igor
Aleiner for many enlightening and helpful discussions. DRR would like
to thank Romain Vasseur for a useful correspondence. This work was
supported by National Science Foundation Grant No. CHE-1464802. DMRG
calculations were performed using the ITensor library, http://itensor.org.
\end{acknowledgments}

\bibliographystyle{apsrev4-1}
\bibliography{/home/yevgeny/Dropbox/Research/Latex/papers/Bibs/library,local}

\clearpage{}

\setcounter{equation}{0} \setcounter{figure}{0} \setcounter{table}{0} \setcounter{page}{1} \makeatletter \renewcommand{\theequation}{S\arabic{equation}} \renewcommand{\thefigure}{S\arabic{figure}} \renewcommand{\bibnumfmt}[1]{[S#1]} \renewcommand{\citenumfont}[1]{S#1}

\section{Supplementary materials}

Here we elaborate on the symmetries of the studied models, derive
the effective Hamiltonian, as well as motivate the chosen disordered
interaction strength.

\subsection{Symmetries}

Even with spatially dependent interactions the Hamiltonian,
\begin{equation}
\hat{H}=-t\sum_{i\sigma}\left(\hat{c}_{i\sigma}^{\dagger}\hat{c}_{i+1,\sigma}+\hat{c}_{i+1,\sigma}^{\dagger}\hat{c}_{i,\sigma}\right)+\sum_{i}U_{i}\hat{n}_{\uparrow i}\hat{n}_{\downarrow i},\label{eq:nonsymm}
\end{equation}
has a $SU\left(2\right)$ symmetry with respect to the rotation of
the spin, with the following generators,
\begin{eqnarray}
\hat{S}^{z} & = & \frac{1}{2}\sum_{i}\left(\hat{n}_{i\uparrow}-\hat{n}_{i\downarrow}\right)\nonumber \\
\hat{S}^{+} & = & \sum_{i}\hat{c}_{i\uparrow}^{\dagger}\hat{c}_{i\downarrow}\nonumber \\
\hat{S}^{-} & = & \sum_{i}\hat{c}_{i\downarrow}^{\dagger}\hat{c}_{i\uparrow}.
\end{eqnarray}
 The modified Hamiltonian,
\begin{eqnarray}
\hat{H}' & = & -t\sum_{i\sigma}\left(\hat{c}_{i\sigma}^{\dagger}\hat{c}_{i+1,\sigma}+\hat{c}_{i+1,\sigma}^{\dagger}\hat{c}_{i,\sigma}\right)+\label{eq:symmetric}\\
 & + & \sum_{i}U_{i}\left(\hat{n}_{\uparrow i}-\frac{1}{2}\right)\left(\hat{n}_{\downarrow i}-\frac{1}{2}\right),\nonumber 
\end{eqnarray}
has an additional $SU\left(2\right)$ symmetry in the charge sector,
which follows particle-hole symmetry,
\begin{equation}
\hat{c}_{i\sigma}\to\left(-1\right)^{i}\hat{c}_{i\sigma}^{\dagger}.
\end{equation}
Its symmetry group is therefore $SU\left(2\right)\times SU\left(2\right)$.
An additional symmetry which exists for the (\ref{eq:symmetric})
Hamiltonian in the $\left(N=L,S_{z}=0\right)$ sector is that the
eigenvalues satisfy, $E\longleftrightarrow-E$.

\subsection{Selection of the disordered interaction strength}

\begin{figure}
\begin{centering}
\includegraphics[width=85mm]{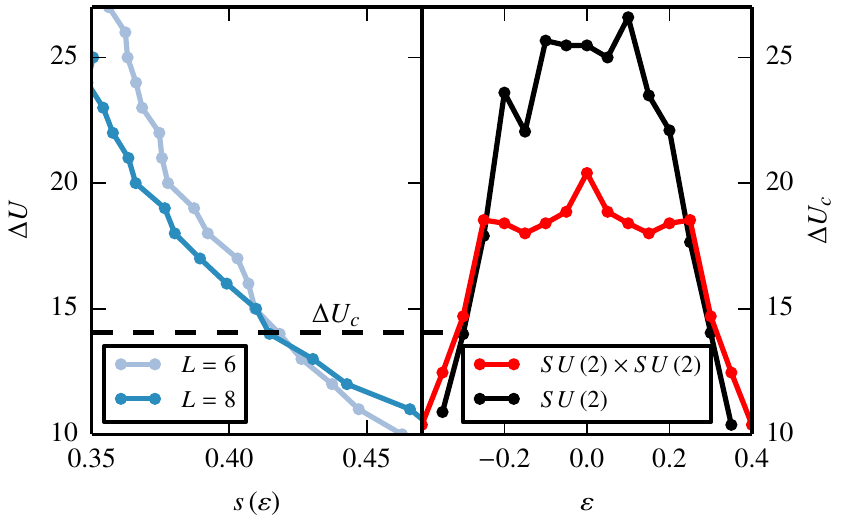}
\par\end{centering}
\caption{\label{fig:ee}\emph{Left panel}: entanglement entropy density $\left(s\left(\varepsilon\right)\right)$
is plotted for various values of interaction disorder for a fixed
energy density, $\varepsilon=-0.3$, and the $SU\left(2\right)$ symmetric
version of model, (\ref{eq:nonsymm}). \emph{Right panel}: the critical
interaction disorder (mobility edge) is plotted versus the energy
density $\varepsilon$ for both for models (\ref{eq:nonsymm}) {[}black{]}
and (\ref{eq:symmetric}) {[}red{]}.}
\end{figure}

To delineate the phase diagram of the system we study the properties
of its eigenstates using exact diagonalization. For this purpose we
calculate the entanglement entropy of all the eigenstates. We first
obtain the reduced density matrix,
\begin{equation}
\rho_{A}\left(E\right)=\text{Tr }_{\bar{A}}\left|\psi\left(E\right)\right\rangle \left\langle \psi\left(E\right)\right|,
\end{equation}
for every eigenstate $\left|\psi\left(E\right)\right\rangle $, where
$A$ is a subsystem (we chose it to be the left half of the system)
and $\bar{A}$ is its complimentary. The entanglement entropy is then
defined as,
\begin{equation}
S\left(E\right)=-\text{Tr }\rho_{A}\left(E\right)\log_{2}\rho_{A}\left(E\right).
\end{equation}
We will designate by $s\left(\varepsilon\right)=S\left(\varepsilon\right)/L$
the entanglement entropy density, and take $\epsilon\equiv\left(E-E_{\min}\right)/\left(E_{\text{max}}-E_{\min}\right)-0.5$,
to be the renormalized unitless energy density, which lies in the
interval $\varepsilon\in\left[-0.5,0.5\right]$. For one-dimensional
systems which we consider in this work, $s\left(\varepsilon\right)$
is expected to be independent of system size if the system is ergodic,
and to be inversely proportional to system size for a non-ergodic
system. The location of the transition could be therefore obtained
by the calculation of $s\left(\varepsilon\right)$ for various disorder
strengths and energy densities. This procedure is presented in Fig.~\ref{fig:ee}.
On the left panel we obtain the location of the mobility edge $\Delta U_{c}$,
for a fixed energy density, $\epsilon$. This is inferred from the
intersection point of plots of $s\left(\varepsilon\right)$ for various
system sizes. Each $s\left(\varepsilon\right)$ is obtained by calculating
$S\left(\varepsilon\right)$ and averaging over a small interval of
energy densities $\delta\varepsilon=0.04$, as also $1000$ disorder
realization. In the left panel of Fig.~\ref{fig:ee} we demonstrate
this procedure for the model (Eq.~\ref{eq:ham_non_sym}) for $\varepsilon=-0.3$.
Due to the fast growth of the Hilbert space $\left(4^{L}\right)$
we have access to only two system sizes, $L=6$ and $8$ in the sector
$N=L,\,S_{z}=0$. After repeating the outlined procedure for a number
of energy densities we obtained the mobility edge, $\Delta U_{c}\left(\varepsilon\right)$,
as a function of the energy density for both symmetric and non-symmetric
versions of the model. As one can see from the right panel of Fig.~\ref{fig:ee},
it has a typical domed shape with states in the middle of the band
less localized than the states with lower (higher) energy densities.

Since in this work we are interested in demonstrating localization
we choose $\Delta U=30$, such that \emph{all} many-state are expected
to be localized for \emph{both} versions of the model.

\subsection{Derivation of the Effective Hamiltonian}

We write the Hamiltonian as
\begin{equation}
\hat{H}=\hat{H}_{K}+\hat{H}_{U},
\end{equation}
where $\hat{H}_{U}$ is the interaction term, 
\begin{equation}
\hat{H}_{U}=\sum_{i}U_{i}\left(\hat{n}_{\uparrow i}-\frac{1}{2}\right)\left(\hat{n}_{\downarrow i}-\frac{1}{2}\right)
\end{equation}
and
\begin{equation}
\hat{H}_{K}\equiv\sum_{i\sigma}t_{ij}\hat{c}_{i\sigma}^{\dagger}\hat{c}_{j\sigma},
\end{equation}
is the perturbation. For $U_{i}<0$ the ground state of $\hat{H}_{U}$
contains only doublons and is $2^{L}$ times degenerate, since a doublon
and a holon have same energies. We define a projector on the zero
and doubly occupied states as, $\hat{P}$ and its complimentary as,
$\hat{Q}=1-\hat{P}$. Then the Hamiltonian can be written in a block
form,
\begin{equation}
\hat{H}=\hat{P}\hat{H}\hat{P}+\hat{Q}\hat{H}\hat{Q}+\left(\hat{P}\hat{H}\hat{Q}+\hat{Q}\hat{H}\hat{P}\right).
\end{equation}
We further divide the Hamiltonian into unperturbed, $\hat{H}_{0}=\hat{P}\hat{H}\hat{P}+\hat{Q}\hat{H}\hat{Q}$
and perturbed parts, $\hat{H}_{1}=\hat{P}\hat{H}\hat{Q}+\hat{Q}\hat{H}\hat{P}$,
and note that $\left[\hat{P},\hat{H}_{0}\right]=0$ but $\left[\hat{P},\hat{H}_{1}\right]\neq0$.
By performing a unitary transformation we eliminate the off-diagonal
term up to second order in $\hat{H}_{1}$. Using Baker-Cambell-Hausdorff
formula we can write,
\begin{eqnarray}
e^{S}\hat{H}e^{-S} & = & \hat{H}_{0}+\left[S,\hat{H}_{0}\right]+\frac{1}{2!}\left[S,\left[S,\hat{H}_{0}\right]\right]+\cdots\\
 & + & \hat{H}_{1}+\left[S,\hat{H}_{1}\right]+\frac{1}{2!}\left[S,\left[S,\hat{H}_{1}\right]\right]+\cdots.\nonumber 
\end{eqnarray}
Setting,
\begin{equation}
\left[S,\hat{H}_{0}\right]=-\hat{H}_{1},\label{eq:sw_req}
\end{equation}
we can recast the series into the form,
\begin{eqnarray}
e^{S}\hat{H}e^{-S} & = & \hat{H}_{0}+\left(\frac{1}{1!}-\frac{1}{2!}\right)\left[S,\hat{H}_{1}\right]\\
 & + & \left(\frac{1}{2!}-\frac{1}{3!}\right)\left[S,\left[S,\hat{H}_{1}\right]\right]+\cdots.\nonumber 
\end{eqnarray}
The effective Hamiltonian to second order in the off-diagonal terms
is therefore given by,
\begin{equation}
\hat{H}_{\text{eff}}=e^{S}\hat{H}e^{-S}=\hat{H}_{0}+\frac{1}{2}\left[S,\hat{H}_{1}\right].
\end{equation}
To calculate $S$ we use (\ref{eq:sw_req}) in the unperturbed basis,
\begin{equation}
\left\langle \alpha\left|\left[S,\hat{H}_{0}\right]\right|\beta\right\rangle =S_{\alpha\beta}\left(E_{\beta}^{\left(0\right)}-E_{\alpha}^{\left(0\right)}\right)=-\left\langle \alpha\left|\hat{H}_{1}\right|\beta\right\rangle ,
\end{equation}
and therefore,
\begin{equation}
S_{\alpha\beta}=\frac{\left\langle \alpha\left|\hat{P}\hat{H}\hat{Q}+\hat{Q}\hat{H}\hat{P}\right|\beta\right\rangle }{E_{\alpha}^{\left(0\right)}-E_{\beta}^{\left(0\right)}}.
\end{equation}
In case that $\hat{P}$ projects to a degenerate subspace, such that
$\hat{H}_{0}\hat{P}\left|\alpha\right\rangle =E_{P}^{\left(0\right)}\hat{P}\left|\alpha\right\rangle $,
we can write,
\begin{equation}
\hat{S}=\hat{P}\hat{H}\left(E_{P}^{\left(0\right)}-\hat{H}_{0}\right)^{-1}\hat{Q}-\hat{Q}\left(E_{P}^{\left(0\right)}-\hat{H}_{0}\right)^{-1}\hat{H}\hat{P}.
\end{equation}
After some algebra we obtain the effective Hamiltonian,
\begin{equation}
\hat{H}_{\text{eff}}=E_{P}^{\left(0\right)}+\sum_{ijkl,\sigma\sigma'}\frac{t_{ij}t_{kl}}{U_{ij}}\hat{P}\hat{c}_{k\sigma'}^{\dagger}\hat{c}_{l\sigma'}\hat{c}_{i\sigma}^{\dagger}\hat{c}_{j\sigma}\hat{P},
\end{equation}
where 
\begin{equation}
U_{ij}\equiv\frac{U_{i}+U_{j}}{2}.
\end{equation}
Finally defining the following pseudo-spin operators,
\begin{equation}
\hat{S}_{i}^{+}=\hat{c}_{i\uparrow}^{\dagger}\hat{c}_{i\downarrow}^{\dagger}\qquad\hat{S}_{i}^{-}=\hat{c}_{i\downarrow}\hat{c}_{i\uparrow},\qquad\hat{S}_{i}^{z}=\frac{1}{2}\left(\hat{n}_{i\downarrow}+\hat{n}_{i\uparrow}-1\right),
\end{equation}
and using the properties of the projectors we obtain the anti-ferromagnetic
random Heisenberg model,
\begin{equation}
\hat{H}_{\text{eff}}=E_{P}^{\left(0\right)}+\sum_{ij}\frac{2\left|t_{ij}\right|^{2}}{\left|U_{ij}\right|}\left(\hat{\mathbf{S}}_{i}\cdot\hat{\mathbf{S}}_{j}-\frac{1}{4}\right).
\end{equation}

\subsection{Finite time and finite size effects}

As in any numerical study, our results are limited to finite sizes
and times. It is therefore pertinent to present evidence that the
observed behavior persists also for larger systems and longer times.
We would like to consider three possible objections to the observed
localization, all related to the relatively high $U$ we use in our
simulation:
\begin{enumerate}
\item For small systems and large interaction, $U/t_{h}\gg1$, the many-body
spectrum includes gaps, which effectively makes the system non-ergodic
\cite{Papic2015}.
\item The system size is too small to either break apart the localized doublons,
or to build a large, almost ``classical'' quasi-particle along the
scenario of delocalization introduced in Ref.~\cite{Potter2016}.
\item The time it takes to observe delocalization is beyond the reach of
our simulations.
\end{enumerate}
\begin{figure}
\begin{centering}
\includegraphics{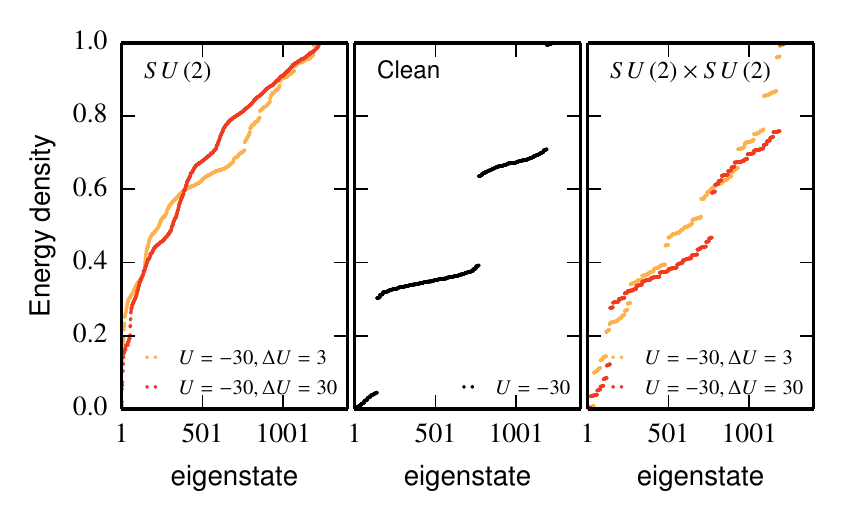}
\par\end{centering}
\caption{\label{fig:spectrum}Eigenvalues of a single disorder realization
normalized to lie in the interval $\left[0,1\right]$. \emph{Left
panel}: $SU\left(2\right)$ symmetric model, (\ref{eq:nonsymm}),
for $\Delta U=3$ (light orange) and $\Delta U=30$ (dark orange).
\emph{Central panel}: clean system, $\Delta U=0$. \emph{Right panel}:
same as left panel but for the $SU\left(2\right)\times SU\left(2\right)$
symmetric model (\ref{eq:symmetric}). For all models $L=7$. }
\end{figure}
We argue that the existence of gaps in the many-body spectrum does
not \emph{necessarily} imply localization, moreover some of the gaps
may very well persist in the thermodynamic limit \cite{Garrison2016}.
In Fig.~\ref{fig:spectrum} we present the many-body eigenvalues
of models (\ref{eq:nonsymm}) and (\ref{eq:symmetric}) for a single
disorder realization, as well as the eigenvalues of the corresponding
clean model $\left(\Delta U=0\right).$ Gaps exist for all three models.
In fact, since for the disordered models we use $-\Delta U\leq U_{i}\leq0$,
gaps are \emph{smaller} for larger $\Delta U$, since disorder enlarges
the phase space of possible $\left(U_{1}+U_{2}+\cdots\right)$ combinations.
Therefore the largest gaps appear for $\Delta U=0$, when only multiples
of $U$ are possible. The transport, on the other hand, is very different
in all three models. It varies from localized, delocalized to subdiffusive,
as we present in the main text. This shows that having gaps in the
many-body spectrum need not have direct implications for ergodicity.

\begin{figure}
\begin{centering}
\includegraphics[width=86mm]{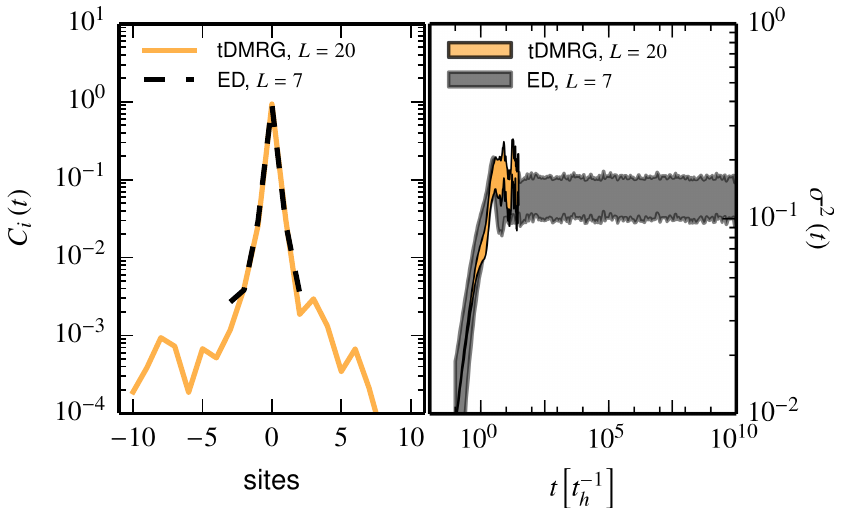}
\par\end{centering}
\caption{\label{fig:su2-no-ones}Charge excitation dynamics for starting from
a random charge configuration \emph{without} singlons calculated using
tDMRG simulation for $L=20$ (orange) and ED for $L=7$ (black/gray).
\emph{Left panel}: Excitation profile at the final times of the simulation
$L=7,$ $t=10^{10}$ and $L=20,$ $t=30$. \emph{Right panel}: charge
excitation width as a function of time on a log-log plot. Shades designate
uncertainty due to averaging. The parameters used are $-30\leq U_{i}\leq0$. }
\end{figure}
The second point concerns the finite size of the system. For small
systems, some excitations might never decay. For example, due to energy
conservation, a doublon with an energy of $U$ would \emph{never}
decay in a system of $N$ particles, if the maximal change of the
single particles is $\delta_{\max}$, such that $U>\delta_{max}N$.
For larger systems a collective (though very rare) excitation of the
particles would be able to break the doublon apart. However since
this process requires a delicate cooperation of the particles the
decay rate of the doublon would be exponentially slow \cite{Sensarma2010}.
Since in the model we study (\ref{eq:ham_non_sym}) a finite doublon
density is necessary to localize the singlons, one may question: \textbf{a)}
if localization might be lost due to doublon decay \textbf{b)} the
small system size prevents the decay to occur. We argue that a \emph{finite}
doublon density will never decay to zero for sufficiently strong interaction
even for \emph{infinite} systems. This means that the mechanism of
localization due to doublons may exist also in the thermodynamic limit,
though with (slightly) renormalized disorder strength. To show that
finite doublon density cannot decay, we repeat the above argument
for a finite system with a minor modification. By duplicating the
small system, and using energy conservation, it is easy to see that
the maximal change in the doublon density is $\delta n_{d}=O\left(1/U\right),$
which means that for sufficiently large $U$ a \emph{finite} density
of doublons cannot decay completely even in an \emph{infinite} system.
Moreover a finite doublon density implies that the temperature is
$T\approx U$, for which the processes required for the decay of the
doublons are not rare anymore, and occur at the rate of $t_{h}.$
The change of the effective disordered potential as a result of the
doublon decay is about $O\left(1/U\right)$. We thus claim that the
effect of the decay of the doublon density on localization is negligible
for large $U$.

In Ref.~\cite{Potter2016} it is suggested that for $SU\left(2\right)$
symmetric models the local degrees of freedom will join into large
``quasi-classical'' objects which will induce delocalization. One
might therefore suspect that either the system is too small to create
such a large object, or alternatively the creation and motion of such
objects might occur at later times. This study however does not specify
what are the required length and time scales to observe delocalization
in such models.

\begin{figure}
\begin{centering}
\includegraphics[width=86mm]{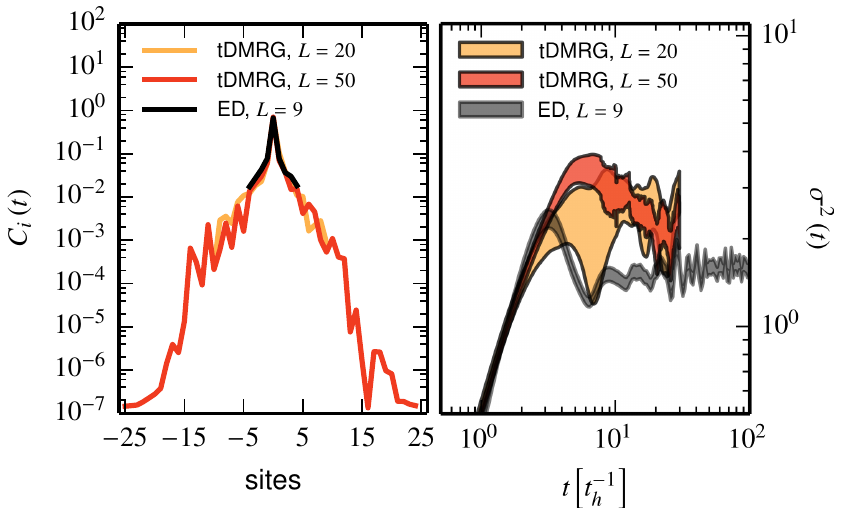}
\par\end{centering}
\caption{\label{fig:su2-all}Charge excitation dynamics for starting from a
random charge configuration \emph{(no constraints)} calculated using
tDMRG simulation for $L=20$ and $50$ (light and dark orange) and
ED for $L=9$ (black/gray). \emph{Left panel}: Excitation profile
at the final times of the simulation $L=9,$ $t=100$ and $L=20,50$,
$t=30$. \emph{Right panel}: charge excitation width as a function
of time on a log-log plot. Shades designate uncertainty due to averaging.
The parameters used are $-30\leq U_{i}\leq0$. }
\end{figure}

In our work we calculate the spread of a \emph{local} excitation which
is described by the correlation function,
\begin{equation}
C_{i}\left(t\right)=\frac{1}{\text{Tr }\hat{P}}\text{Tr }\hat{P}\left(\hat{n}_{i}\left(t\right)-1\right)\left(\hat{n}_{0}-1\right).
\end{equation}
The locality of the operators which appear in the correlation function,
as well as the locality of the Hamiltonian (\ref{eq:ham_non_sym}),
guarantee
\begin{equation}
\left\Vert \left\langle \hat{O}_{i}\left(t\right)\hat{O}_{j}\right\rangle \right\Vert \leq\exp\left[-a\left(\left|i-j\right|-vt\right)\right],
\end{equation}
\begin{figure}
\begin{centering}
\includegraphics[width=86mm]{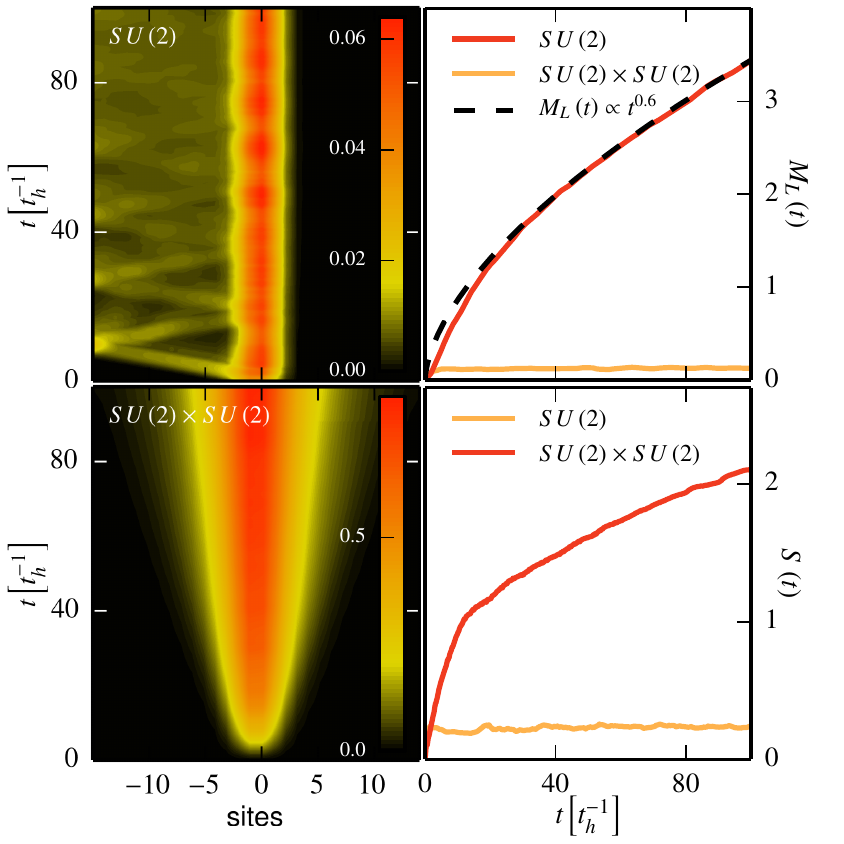}
\par\end{centering}
\caption{\label{fig:domain_wall} On the left, the charge density perturbation,
$\left\langle \hat{n}_{i}\left(t\right)-\hat{n}_{i}\left(0\right)\right\rangle $,
is plotted vs time and space for models (\ref{eq:nonsymm}) {[}top{]}
and (\ref{eq:symmetric}) {[}bottom{]}. On the right the total charge
on the LHS, $M_{L}\left(t\right),$ is plotted versus time on the
top panel and the entanglement entropy, $S\left(t\right)$ versus
time is plotted on the bottom. The data has been obtained using tDMRG
for $L=30$ and $\Delta U=30$, and averaged over at least 300 disorder
realizations.}
\end{figure}
following from rigorous Lieb-Robinson bounds \cite{Lieb1972}, where
$v$ is the maximal spread velocity, which is of order of the hopping
rate (here $O\left(1\right)$). This means that for any \emph{finite}
time, the excitation is bounded by an exponential in the distance
$\left|i-j\right|$, namely it is effectively contained in a finite
box. \emph{Finite size} effects could be therefore eliminated to any
required precision \emph{up to any finite time}. For localized systems
the bound is uniform in time, which signifies the absence of transport
\cite{Friesdorf2015}. In our work, by looking at the profile of the
excitation as well as its width, we show that for the $SU\left(2\right)$
symmetric model (\ref{eq:nonsymm}) this bound is indeed uniform at
least up to some finite time. As can be seen from the left panel of
Fig.~\ref{fig:su2-no-ones}, starting from an initial configuration
of only doublons, the $L=7$ system and the $L=20$ system essentially
give the same results, while compared at strikingly different times
$t=30$ $\left(L=20\right)$ versus $t=10^{10}$ $\left(L=7\right)$.
This signifies both absence of charge transport, and effective elimination
of finite size effects. For \emph{unconstrained} random initial configuration,
as can be seen from Fig.~\ref{fig:su2-all}, non-negligible residual
tails of the excitation prevent elimination of finite size effects
for a system size of $L=9$ (ED). Nevertheless, the excitation profile
stays almost the same across increasing system sizes of $L=9,$ $20$
and $50$, and the width of the excitation appears to be converged
already for $L=20$. The tDMRG simulation demands considerable resources
in this regime due to the initial fast growth of the entanglement
entropy, which results from ballistic motion of the singlons between
the doublons. This results in poor averaging (100-300 realizations)
and a bias towards more localized charge configurations. Notwithstanding,
the appearance of localization plateau, while less convincing compared
to ``doublons only'' case, is evident here as well.

\subsection{Domain wall initial condition}

We have also considered the dynamics starting from a domain wall initial
conditions. This type of initial conditions is more amenable to experimental
study in cold atoms \cite{Choi2016}. We have chosen an initial condition
such that the right half of the system is completely full, while the
left half is empty, and measured the total number of particles on
the LHS, $M_{L}\left(t\right).$ In Fig.~\ref{fig:domain_wall} one
can see the profile on the charge density perturbation,$\left\langle \hat{n}_{i}\left(t\right)-\hat{n}_{i}\left(0\right)\right\rangle ,$
on the left, as also the charge transfer, $M_{L}\left(t\right)$ and
$S\left(t\right)$ on the right. For the $SU\left(2\right)$ symmetric
model described by (\ref{eq:nonsymm}) the ballistic jet of the singlons
is clearly visible, as is its reflections from the simulation boundaries
as well. The core of the system however remains localized. The $SU\left(2\right)\times SU\left(2\right)$
symmetric model described by (\ref{eq:symmetric}) is clearly delocalized
and on the accessible time scales exhibits transport very close to
diffusive, 
\begin{equation}
M_{L}\left(t\right)\propto t^{0.6},
\end{equation}
with entanglement entropy which is almost linearly growing with time.

\end{document}